\documentclass[12pt,a4paper]{article}

\newlength{\abstwidth}
\usepackage{graphicx}

\renewcommand{\b}{\mathrm{b}}
\renewcommand{\c}{\mathrm{c}}

\newcommand{\e}{\mathrm{e}}

\newcommand{\g}{\mathrm{g}}

\newcommand{\p}{\mathrm{p}}
\newcommand{\q}{\mathrm{q}}

\renewcommand{\t}{\mathrm{t}}
\renewcommand{\u}{\mathrm{u}}

\renewcommand{\H}{\mathrm{H}}

\newcommand{\W}{\mathrm{W}}

\newcommand{\Z}{\mathrm{Z}}
\newcommand{\bbar}{\overline{\mathrm{b}}}
\newcommand{\cbar}{\overline{\mathrm{c}}}

\newcommand{\pbar}{\overline{\mathrm{p}}}
\newcommand{\qbar}{\overline{\mathrm{q}}}

\newcommand{\tbar}{\overline{\mathrm{t}}}
\newcommand{\ubar}{\overline{\mathrm{u}}}

\newcommand{\ee}{\e^+\e^-}

\newcommand{\pp}{\p\p}
\newcommand{\ppbar}{\p\pbar}
\newcommand{\alphas}{\alpha_{\mathrm{s}}}

\newcommand{\pT}{p_{\perp}}

\begin{document}
\sloppy

\pagestyle{empty}
\begin{flushright}
LU TP 19-31\\
MCnet-19-16\\
July 2019
\end{flushright}

\vspace{\fill}
\begin{center}

{\Huge\bf The PYTHIA Event Generator:}\\[4mm]
{\Huge\bf Past, Present and Future%
\footnote{submitted to 50 Years of Computer Physics Communications: 
a special issue focused on computational science software}}\\[10mm]
{\Large Torbj\"orn Sj\"ostrand}\\[3mm]
{\it Theoretical Particle Physics,}\\[1mm]
{\it Department of Astronomy and Theoretical Physics,}\\[1mm]
{\it Lund University,}\\[1mm]
{\it S\"olvegatan 14A,}\\[1mm]
{\it SE-223 62 Lund, Sweden}
\end{center}

\vspace{\fill}

\begin{center}
\begin{minipage}{\abstwidth}
\begin{center} {\bf Abstract}
\end{center}
The evolution of the widely-used \textsc{Pythia} particle physics 
event generator is outlined, from the early days to the current status 
and plans. The key decisions and the development of the major physics 
components are put in context.
\end{minipage}
\end{center}

\vspace{\fill}
\phantom{dummy}

\clearpage

\pagestyle{plain}
\setcounter{page}{1}

\section{Introduction}
\label{intro}

The \textsc{Pythia} event generator is one of the most commonly 
used pieces of software in particle physics and related areas, 
either on its own or ``under the hood'' of a multitude of other programs. 
The program is designed to simulate the physics processes that can 
occur in collisions between high-energy particles, e.g.\ at the 
LHC collider at CERN. Monte Carlo methods are used to represent
the quantum mechanical variability that can give rise to wildly 
different multiparticle final states under fixed simple initial 
conditions. A combination of perturbative results and models for 
semihard and soft physics --- many of them developed in the 
\textsc{Pythia} context --- are combined to trace the evolution 
towards complex final states. 
 
\begin{table}[tp]
\caption{The main versions of \textsc{Jetset} and \textsc{Pythia}, 
with their date of appearance and published manuals where relevant. 
\protect\label{JetsetPythiaver}} 
\begin{center}
\begin{tabular}[t]{|c|c|c|}
\hline
\multicolumn{3}{|c|}{\textsc{Jetset} versions}\\
\hline
No. & Date  & Publ. \\[1mm]
\hline
1   & Nov 78 & \cite{Sjostrand:1978dj} \\
2   & May 79 & \cite{Sjostrand:1979vc} \\
3.1 & Aug 79 &   ---  \\
3.2 & Apr 80 & \cite{Sjostrand:1980ka} \\
3.3 & Aug 80 &   ---  \\
4.1 & Apr 81 &   ---  \\
4.2 & Nov 81 &   ---  \\
4.3 G & Mar 82 & \cite{Sjostrand:1982fn} \\
4.3 E & Jul 82 & \cite{Sjostrand:1982am} \\
5.1 & Apr 83 &   ---  \\
5.2 & Nov 83 &   ---  \\
5.3 & May 84 &   ---  \\
6.1 & Jan 85 &   ---  \\
6.2 & Oct 85 & \cite{Sjostrand:1985ys} \\
6.3 & Oct 86 & \cite{Sjostrand:1986hx} \\
7.1 & Feb 89 &   ---  \\
7.2 & Nov 89 &   ---  \\
7.3 & May 90 & \cite{Sjostrand:1992mh} \\
7.4 & Dec 93 & \cite{Sjostrand:1993yb} \\[1mm] 
\hline
\end{tabular}
\begin{tabular}[t]{|c|c|c|}
\hline
\multicolumn{3}{|c|}{\textsc{Pythia} versions}\\
\hline
No. & Date  & Publ.   \\[1mm]
\hline
1   & Dec 82 & \cite{Bengtsson:1982jr} \\
2   &  ---  &   \\
3.1 &  ---  &   \\
3.2 &  ---  &   \\
3.3 & Feb 84 & \cite{Bengtsson:1984yx}  \\
3.4 & Sep 84 & \cite{Bengtsson:1984yx}  \\
4.1 & Dec 84 &  \\
4.2 & Jun 85 &  \\
4.3 & Aug 85 &  \\
4.4 & Nov 85 &  \\
4.5 & Jan 86 &  \\
4.6 & May 86 &  \\
4.7 & May 86 &  \\
4.8 & Jan 87 & \cite{Bengtsson:1987kr} \\
4.9 & May 87 &  \\
5.1 & May 87 &  \\
5.2 & Jun 87 &  \\
5.3 & Oct 89 &  \\
5.4 & Jun 90 &  \\
5.5 & Jan 91 &  \\
5.6 & Sep 91 & \cite{Sjostrand:1992mh} \\
5.7 & Dec 93 & \cite{Sjostrand:1993yb} \\
6.1 & Mar 97 & \cite{Sjostrand:2000wi} \\
6.2 & Aug 01 & \cite{Sjostrand:2001yu} \\
6.3 & Aug 03 & \cite{Sjostrand:2003wg} \\
6.4 & Mar 06 & \cite{Sjostrand:2006za} \\
7.0 & May 00 & \cite{Bertini:2000uh}\\
8.1 & Oct 07 & \cite{Sjostrand:2007gs}  \\
8.2 & Oct 14 & \cite{Sjostrand:2014zea} \\[1mm]
\hline
\end{tabular}
\end{center}
\end{table}

The program roots stretch back over forty years to the \textsc{Jetset} 
program, with which it later was fused. Twelve manuals for the 
two programs have been published in Computer Physics Communications
(CPC), Table~\ref{JetsetPythiaver}, together collecting over 20\,000 
citations in the Inspire database. A longer (580~pp) physics 
description and manual, published in JHEP, is one of only six 
Inspire entries to exceed 10\,000 citations. This was an exception,
however, and JHEP immediately recognized its ``mistake'' by 
thenceforth banning all publication of manuals. Thus CPC has remained
the premier place where descriptions of high energy physics programs
has always been welcome, which deserves to be recognized at this
50$^{\mathrm{th}}$ anniversary of CPC. 

In the present contribution to the celebration, I will outline the 
historical evolution of \textsc{Pythia/Jetset}, its current status 
and plans for the future. The description is anecdotal, with emphasis 
on some of the early decisions and key concepts that came to shape the 
continued evolution of the program(s). Therefore it should not be viewed 
as a full-scale review of the \textsc{Pythia} physics and code, 
let alone of the broader fields of particle physics phenomenology 
and event generators. The text is also not strictly chronological, 
but typically presents a topic at a time when it became important, 
with brief comments on earlier roots and later developments. Finally,
since so many diverse topics enter in passing, the bibliography 
had to be restrictive. Further references can be found in the articles 
quoted, notably in the \textsc{Pythia}~6.4 manual \cite{Sjostrand:2006za}
and in some reviews \cite{Andersson:1983ia,Buckley:2011ms,Sjostrand:2017cdm}.

\section{In the beginning}
\label{beginning}

The first code of what was to become \textsc{Jetset} was written 
in May 1978. It sprung out of studies in Lund on the structure of 
the fragmentation of a single quark into a jet of hadrons, that had 
begun about a year earlier, led by Bo Andersson and G\"osta Gustafson
\cite{Andersson:1977qx,Andersson:1978vj}. These early studies 
were based on a recursive structure, expressible in terms of an 
integral equation that could be solved analytically in simple cases. 
The article by Field and Feynman \cite{Field:1977fa} introduced a 
similar model, but also took the step of simulating the process  on a 
computer with Monte Carlo techniques. Thereby each quark jet became 
associated  with an explicit list of particles, opening up for more 
detailed studies than is possible analytically. At the time it was 
quite a novel idea to most people. Actually, Artru and Mennessier 
had introduced and simulated a very interesting fragmentation model 
some years earlier \cite{Artru:1974hr}, for a kicked-apart
quark--antiquark pair connected by a linear-potential force-field 
``string''. With a constant probability per unit space--time area for 
the string to break, a continuous hadron mass spectrum is obtained.
At the time this was viewed as a limitation, but today it could be 
considered as the blueprint for a Lorentz covariant cluster fragmentation 
model. Unfortunately only few people were aware of that article in 1978.      

By the suggestion of the head of the Lund group, Bengt E. Y. Svensson,
two young PhD students, Bo S\"oderberg (who coined the \textsc{Jetset}
name) and myself, were tasked with reproducing the work of Field and 
Feynman, and extending it to the analytical model developed in Lund. 
The practical conditions were not the best. There only existed one 
computer at Lund University with the capacity to run such programs, and 
it had a clock speed of approximately 1~MHz and a CPU memory of 1~MB. 
Worse, input was by a stack of punched cards, where each card 
corresponded to one 80-character line of Fortran~77 code. The card 
reader had a tendency to fail --- at worst meaning a destroyed card 
--- every few hundred cards, thereby favouring compact programming. 
Output was by line printer, some unpredictable 10 to 30 minutes later. 
The smallest error and you had to retype the affected card(s), reread 
the whole program and wait another 20 minutes. The Fortran language 
was new to us, but easy to learn. The CERN \textsc{Hbook} histogramming 
package \cite{Brun:1977wm} was one of the few libraries available, and 
its user interface influenced how we thought about having a library of 
physics and service methods operating on a common event record.

It is maybe not surprising that the seniors, especially Bo Andersson, 
initially were quite hesitant whether there was any future in doing 
physics studies this way. Or that Bo S\"oderberg grew tired of it, 
leaving me to continue alone. After about a year conditions slowly 
started to improve, in particular with the introduction of simple 
(phone-line) terminals and the ability to edit and save programs on 
the computer.  

While the fate of the nascent \textsc{Jetset} program still hang 
in the balance, the PETRA $\ee$ collider at DESY started to 
produce results on jets, and notably found three-jet events 
\cite{Brandelik:1979bd}, offering evidence for the existence of gluons. 
The experimental observations were backed up by comparisons with the 
event generator developed by Hoyer et~al.~\cite{Hoyer:1979ta} and, 
later,  with the one by Ali et~al.~\cite{Ali:1979wj}. Both of these were 
based on the concept of independent fragmentation, wherein each of the 
$\q$, $\qbar$ and $\g$ jets are assumed to fragment symmetrically
around a jet axis defined by the direction of the respective parton
in the CM frame of the event. In Lund, instead another picture had 
been developed, string fragmentation. Here the confining colour field 
is approximated by a massless relativistic string, with 
gluons represented by pointlike momentum-carrying ``kinks'' 
\cite{Andersson:1979ij}. For a $\q\qbar\g$ event a colour field is 
then stretched from the $\q$ via the $\g$ and on to the $\qbar$, 
with no string directly connecting the $\q$ and $\qbar$. This leads to 
the latter interjet region  being depleted of particle production, 
whereas the other two are enhanced.

A simple code to generate events according to this approach was written, 
and we studied how our model would affect event properties, notably
the structure of the gluon jet \cite{Andersson:1980vk}. Our interactions 
with the PETRA experimentalists led to the JADE collaboration being able 
to present first evidence for the string picture at the Moriond meeting
in March 1980 \cite{Bartel:1980vf}, and this was followed by further studies 
supporting the string picture \cite{Bartel:1981kh}. This breakthrough 
overcame the misgivings the seniors had had previously, and from then on 
event generator development came to play an increasing role in the Lund 
group activities. The string model was not generally accepted initially, 
however. One of the reasons was that the TASSO collaboration failed to 
reproduce the JADE results (unpublished, but widely known). It was only 
in 1982, when the CELLO collaboration noted that string fragmentation 
requires a bigger $\alphas$ than independent fragmentation to describe 
the same three-jet rate \cite{Behrend:1982yv}, that it became important 
for DESY to see the issue resolved. A new TASSO analysis effort then 
found a mistake in the previous one, that had masked effects in the data, 
and matters began to settle down \cite{Althoff:1985wt}.

Even before this happened, another development was that the 
newly-formed LEP collaborations mainly opted for \textsc{Jetset} 
in their QCD studies, rather than for the Hoyer et~al.\ and Ali et~al.\
programs. (And similarly at PEP and TRISTAN.)
A reason was that the latter two programs had been written on the
DESY mainframe, making extensive but inefficient use of existing 
DESY software, such that they did not fit in the smaller CPU memory 
of the CERN mainframe at the time. By contrast, \textsc{Jetset} was
written from scratch by one person, in a more efficient and 
compact manner, and easily fitted. This experience came to set 
its mark on further \textsc{Jetset/Pythia} development, in that an 
ambition remained for programs to be designed to run standalone 
and have a modest footprint. Thereby the programs early on could be 
installed across the World, including places with modest computing 
support, and also rapidly could make its way onto personal computers 
when later these came along. Another relevant factor was that Lund was 
a remote isolated place in the days before Internet, making the effort 
spent on writing detailed manuals essential for successful usage 
elsewhere.

In Lund, there were now two lines of development. One was to make the 
string fragmentation model itself more sophisticated. This involved a 
number of topics, concerning flavour composition and in particular 
baryon production, the transverse broadening by nonperturbative and
semiperturbative mechanisms, and the longitudinal sharing of momenta
\cite{Andersson:1983jt}.
The Physics Reports written in 1982 \cite{Andersson:1983ia} fairly well 
summarizes this work, and marks the end of the most intense period 
of string fragmentation development. But further studies have been made 
now and then over the years, some incorporated into the public code 
and others not. Of special practical importance is the extension to 
string topologies with arbitrarily many gluons between the quark and 
antiquark ends of a string \cite{Sjostrand:1984ic}, and the extension 
to Y-shaped ``junction'' topologies \cite{Sjostrand:2002ip},
where three strings meet, as can be the case if the three valence 
quarks of a proton are kicked out in different directions. 

The other line was the extension of the $\ee$ machinery to other
collision process types. The first of these came from a request/plea 
of the spokesperson of the European Muon Collaboration to provide 
a simulation of Deeply Inelastic Scattering. Thus the \textsc{Lepto}
program \cite{Ingelman:1980qw} was born, in the first instance coded 
by me but soon taken over by co-student Gunnar Ingelman. It included 
NLO (next-to-leading order) corrections, i.e.\ combined the 
$\ell \q \to \ell \q$ processes with $\ell \q \to \ell \q \g$ and 
$\ell \g \to \ell \q \qbar$. 

\textsc{Lepto} introduced a pattern that was to be repeated in the 
years to come: a process-specific code that uses (or not) perturbation 
theory to set up parton-level configurations with specified colour-flow 
string connections, followed by a call to the \textsc{Jetset} string 
fragmentation methods to handle the rest of the story. The 
\textsc{Jetset} code itself consisted of two parts, one being the 
dedicated setup of perturbative $\ee$ annihilation processes, the other 
the multipurpose fragmentation methods. (In my first contacts with 
Computer Physics Communications, in 1982, these two parts therefore 
were submitted separately, according to the author instructions of 
the time. Submission was by snail mail, and code was sent in on
storage tapes.) 

The next student to join was Hans-Uno Bengtsson, who began to study
string effects in hadron collisions, i.e. how the colour
connections between partons lead to some regions having an enhanced
particle flow and others a depleted one. The program \textsc{Compton}
considered topologies where a photon recoils against a quark or a gluon, 
whereas \textsc{High-}$\pT$ addressed general $2 \to 2$ scatterings of
partons. The latter in particular raised a completely new issue,
namely that different colour flow topologies could interfere with each 
other, giving rise to a (positive or negative) fraction of the cross 
section that could not be associated with a well-defined colour 
topology. Hans-Uno used the different pole structures to find a 
sensible but not unique subdivision. 

The multiplication of program names was a concern that Hans-Uno and I 
discussed and, both being avid readers of \textit{The Histories} 
by Herodotus of Halicarnassus, we settled on the name \textsc{Pythia}
for the combined framework for $\p\p/\p\pbar$ collisions.

\section{The formative years}
\label{formative}

In early 1983 Gunnar left for CERN and I for DESY, while Hans-Uno
a year later moved to UCLA. During my time at DESY I was busy
with extending string fragmentation to arbitrarily long colour chains,
implementing second-order matrix elements (MEs) in $\ee$ annihilation,
studying the relationship between hadronization models and $\alphas$
determinations, and supporting the early HERA experimental studies, 
wherein \textsc{Lepto} came to be used, extended by Gunnar from QED 
to full electroweak boson exchange.

Then I moved to Fermilab just in time to go to the 1984 Snowmass
meeting, where future SSC physics was intensely studied and discussed.
The main two event generators were \textsc{Isajet} \cite{Paige:1985tj}
and \textsc{Fieldajet} \cite{Field:1984yz}, where the former was publicly 
available and the latter not, but both with lots of results to show, 
and I learned from Frank Paige and Rick Field about them. Both programs 
were based on independent fragmentation, which in principle gave 
\textsc{Pythia} an edge. But this was moot, since \textsc{Pythia} 
lacked the parton showers that gave the other two programs realistic 
jet shapes. There were also other shortcomings that made \textsc{Pythia} 
unsuited for collisions at high energies.  

By agreement with Hans-Uno, therefore I began to work towards making
\textsc{Pythia} competitive. (Technology had progressed; in the autumn of 
1984 it became possible to send e-mail and thereby code between Fermilab 
and UCLA, opening up for long-distance collaboration. Europe was still 
out of reach.) A first step was to include final-state parton showers, 
which was done by coding up two already existing algorithms, 
a ``conventional'' shower proposed by Kajantie and Pietarinen 
\cite{Kajantie:1980we} and the coherent one of Marchesini and Webber
\cite{Marchesini:1983bm}. Since they could also be used for $\ee$ studies, 
they were made part of \textsc{Jetset}. 

The challenge, however, was initial-state showers. The few studies that 
had been made at that point, notably with COJETS \cite{Odorico:1983yf}, 
were based on forwards evolution, wherein the cascade was started at a 
low $Q_0$ scale and then traced towards larger $Q$ scales. This means 
that the hard interaction is not predetermined, which in practice leads 
to low efficiencies. At Snowmass I had been asking parton distribution 
function (PDF) experts like Wu-Ki Tung about the prospects of 
backwards PDF evolution, from large to small $Q$ scales, i.e.\ in some 
sense backwards in time, with the conclusion that no appropriate tools 
existed. The key point is that, even if the forwards-evolution splitting 
kernels are flavour-symmetric, the flavour content of the proton is not, 
and this must be reflected in the backwards evolution. My proposed 
solution was an algorithm where the PDF evolution equations are inverted 
to give a backwards evolution not only containing splitting kernels, 
like in forwards evolution, but also ratios of PDFs \cite{Sjostrand:1985xi}. 
This approach still is a key component of \textsc{Pythia}, and is used 
in most other current generators.

One hope had been that the introduction of ISR and FSR should make it 
possible to describe data coming from the S$\p\pbar$S. This was not the 
case; notably the long tail towards large multiplicities in UA5 data 
\cite{Alner:1984is}was absent in the generator, as were the strong UA5 
long-distance forward--backward multiplicity correlations 
\cite{Ansorge:1988fg}. Frank Paige had explained
the multi-Pomeron-based model in \textsc{Isajet}, a purely soft approach
also used elsewhere. On the other hand, there had also been interesting
studies on Double Parton Scattering (DPS), wherein two parton--parton 
collisions occur in the same hadron--hadron one, but mainly viewed as a rare 
high-$\pT$ process \cite{Landshoff:1978fq,Goebel:1979mi,Paver:1982yp,%
Humpert:1983pw}. Finally, UA1 had found minijets down to $\pT = 5$~GeV 
\cite{Ceradini:1985vm}, a practical rather than a physics cutoff.
These considerations I brought together in a first model for 
MultiParton Interactions (MPIs) \cite{Sjostrand:1985vv}, wherein 
parton--parton collisions are assumed to be perturbatively calculable
down to a tunable scale $p_{\perp\mathrm{min}}$, at the time 1.6~GeV. 
(To be changed somewhat over the years, e.g. as a function of the 
small-$x$ behaviour of PDFs.) A variable number of MPIs is obtained 
by evolution downwards in $\pT$ from the maximal scale, and this 
variability is what is needed to describe the above-mentioned UA5 data. 
The single-interaction and DPS frameworks are recovered at large 
$\pT$ scales, while most MPIs occur at smaller $\pT$ values and then 
are not too dissimilar from the $\pT = 0$ soft Pomerons. 

A key issue was how the colour flow between different MPIs is related, 
and this led to a picture where colour reconnection (CR) was allowed.
Over the years to come, this MPI+CR model was to be made successively more
sophisticated \cite{Sjostrand:2017cdm}. It may be the \textsc{Pythia} 
core component on which most effort has been spent. While initially met with 
considerable scepticism, today it is generally accepted and the approach 
has spread to other generators. 

The third \textsc{Pythia} improvement, in addition to showers and MPIs, 
was that we added many new processes. Initially the program had only
contained QCD and QED $2 \to 2$ processes, but now $\Z^0$, $\W^{\pm}$
and $\H^0$ production were added, singly, in pairs or together with 
a parton. A first few Beyond the Standard Model (BSM) processes were
also added. All cross sections and differential decay 
distributions had to be coded by hand from formulae in the literature,
sometimes with issues that needed to be sorted out with the
respective authors, which made many new processes rather time-consuming 
to implement. In addition to the explicit hard processes, the MPI 
machinery allowed an inclusive description of inelastic nondiffractive 
events, to which elastic and very simple diffractive events could be
added to obtain a description of all components of the total cross 
section.

By this evolution \textsc{Pythia} became competitive with the other 
generators on the market, and in some respects surpassed them. 
At Snowmass 1986 Hans-Uno and I could fully join and contribute to 
the SSC physics studies, such as Higgs searches. Getting to be used 
by the big experimental collaborations was a more gradual process.
The most important step was the year-long 1990 ``Aachen'' workshop
that marked the beginning in earnest of physics and detector studies 
for the LHC. Being at CERN in 1989--95, I got involved 
in just about all the different physics subgroups. Thus most of 
my time was spent to explain various physics aspects and to cater to 
the generator needs in the subgroups, e.g.\ by implementing 
new BSM processes, and that meant many new \textsc{Pythia} users. 
When the LHC detectors gradually were designed, \textsc{Pythia}
thereby came to be the main generator used to study the performance
under different assumptions, and this propagated on through the
subsequent physics preparations and into the operations era. 

LEP started running in 1989, so also preparations for that took 
some effort in the second half of the eighties. It was clear that 
parton showers would play a key role 
in order to produce multijet final states, but also that three-jet
events would be a main staple of QCD studies. The Kajantie--Pietarinen 
algorithm did not contain coherence, and the Marchesini--Webber one 
did not cover the full three-jet phase space. Instead a somewhat simpler 
shower was developed \cite{Bengtsson:1986hr}. It was a hybrid, involving 
evolution in $m^2$ but with angular-ordering cuts to ensure coherence, 
and as such with limitations. It had one redeeming feature for its day, 
however, in that it covered the full three-body phase space at a 
calculable rate that was slightly above the ME one, neglecting Sudakov 
factor \cite{Sudakov:1954sw} effects. The veto algorithm 
\cite{Sjostrand:2006za} could therefore be used to reduce 
the first emission rate down to the ME level, multiplied by a Sudakov 
that is fixed by the ME and the choice of evolution variable. The same 
``ME corrections'' or ``ME exponentiation'' formalism has later been 
applied also to $\pT$-ordered showers. It has been extended to cover 
the MEs for essentially all SM and SUSY two-body decays that are followed 
by a gluon emission \cite{Norrbin:2000uu}, which means that the default 
treatment of $\gamma^* / \Z^0$, $\W^{\pm}$, $\t$ and $\H^0$ decays is 
accurate to NLO. The same approach can also be used to exponentiate the 
MEs for single $\gamma^* / \Z^0 / \W^{\pm} / \H^0$ production in 
association with a quark or gluon \cite{Miu:1998ju}. 

At the same time another shower was being developed, also in Lund
(where I was 1985--89). The Leningrad group had found that the soft-gluon 
emission pattern around a $\q\qbar\g$ topology could be viewed as a sum 
of radiation off two independent dipoles, $\q\g$ and $\g \qbar$, mimicking 
the nonperturbative string picture \cite{Azimov:1986sf}. G{\"o}sta 
Gustafson had realized that this offered a starting point to formulate a 
shower as a successive branching of dipoles \cite{Gustafson:1986db}, an 
idea that today is a standard choice for most shower algorithms, also the
\textsc{Pythia} ones \cite{Sjostrand:2004ef}, with local variations. 
A student had been put to implement this approach, that was to become 
the \textsc{Ariadne} program, but progress was slow. (Else this approach 
might have found its way into \textsc{Jetset} sooner.) It was only later, 
with a second student, Leif L\"onnblad, that \textsc{Ariadne} took off 
\cite{Lonnblad:1992tz} and usually offered the best shower description 
of LEP data. 

\section{The convergence of PYTHIA and JETSET}
\label{convergence}

In the late eighties Hans-Uno moved away from particle physics. With a 
sole developer/maintainer of both \textsc{Jetset} and \textsc{Pythia}, 
the two programs therefore could be increasingly coordinated. In addition 
the distinction between \textsc{Jetset} for $\ee$ and \textsc{Pythia} 
for $\pp / \ppbar$ began to crumble. For LEP~2 preparations it was 
necessary to implement $\W^+ \W^-$, $\gamma^* / \Z^0 \, \gamma^* / \Z^0$, 
$\H^0 \Z^0$, etc. But these production processes were already available 
for $\pp / \ppbar$, and could trivially be extended to $\ee$. 
The same goes for the key LEP~1 process of $\Z^0$ production and decay.
Therefore \textsc{Pythia} was becoming the prime repository of (hard
and soft) processes, plus ISR and MPI, with \textsc{Jetset} handling 
the subsequent hadronization, plus FSR. The old \textsc{Jetset}
$\ee$ machinery lived on, since it did allow for arbitrary transverse
and longitudinal $\ee$ beam polarization, and contained second-order
matrix elements, but gradually faded away, and is not ported to 
\textsc{Pythia}~8. 

Given the tighter integration, the first combined physics description 
and manual of both programs appeared in 1992, already then 280 pages long. 
It was gradually updated and extended over the years to come, reaching 
480 pages (with same formatting; 580 in JHEP) in the final 
\textsc{Pythia}~6.4 article in 2006. Before then, versions of this 
evolving document only appeared as preprints, with hardcopies in 
steady demand at the CERN computing center. (Already from the 
mid-eighties the center had supplied ad hoc collections of separate 
pieces of documentation.) The size reflects not only the breadth of 
physics covered but also the access given to all methods and parameters. 
Furthermore it is important to offer alternatives in order to test models, 
such that one can establish not only what works but also what does not, 
and this adds to the size.    

Finally, in 1996, the \textsc{Jetset} code was integrated into the 
\textsc{Pythia} package, and program elements renamed to adhere to
\textsc{Pythia} conventions. 

Integration of course was not the only theme, but also continued
evolution and expansion in a number of respects. It would carry too 
far to give a full coverage of the evolution up to the end of  
\textsc{Pythia}~6, but some examples are given below, in no particular
chronological order.
 
One of the prime objectives of LEP~2 was to determine the $\W$ mass.
The fully hadronic decays 
$\e^+ \e^- \to \W^+ \W^- \to \q_1 \qbar_2 \q_3 \qbar_4$ were expected
to give about as small errors as the semileptonic 
$\q_1 \qbar_2 \ell \nu_{\ell}$ ones. One study suggested that 
colour reconnection would completely mess up the hadronic channel 
\cite{Gustafson:1988fs}, but Valery Khoze and I did a more detailed study, 
showing that the perturbative effects should be under control,
whereas nonperturbative CR effects could give a mass uncertainty of 
order 40~MeV \cite{Sjostrand:1993hi}. This was based on two new models 
wherein the space-time string overlap between the two $\W$ systems 
(including parton-shower effects) is traced, assuming similarities 
with flux lines either in type I or in type II superconductors. 

Another limiting factor could be Bose-Einstein (BE) effects, also linking 
the two $\W$ systems. BE is not part of the standard simulation chain 
in Pythia. Like many other aspects of generator physics it is grounded
in quantum mechanics, but it is a truly nonlocal phenomenon that cannot
even to first approximation be reduced to a simple probabilistic 
step-by-step procedure, unlike parton showers or string fragmentation. 
One possible way out is to assign events a weight once the final hadron 
topology is known. In \textsc{Pythia} another approach has been 
implemented, in which the momenta of particles are shifted so as to 
change distributions from approximate phase space to the intended two-body 
correlation function. The shift of each particle is calculated as a 
vector sum of the shifts inside each pair of identical particles that 
the particle belongs to. Leif L{\"o}nnblad and I extended this approach 
to $\W$ pairs, trying a few alternatives, and came to the conclusion 
that $\W$ mass uncertainties from BE could be even somewhat larger than 
those from CR.

Both issues were studied at LEP~2. With predicted effects at the edge of 
detectability, it is maybe not surprising that different conclusions 
were reached. In the final combined analysis \cite{Schael:2013ita} there 
are convincing evidence for a CR rate in agreement with predictions, 
whereas no evidence was found for BE effects.  

The photon is not as simple as it might seem at first glance, since it 
can fluctuate (electromagnetically) into a $\q\qbar$ pair and then 
undergo strong interactions. A low-virtuality fluctuation 
has time to emit further gluons and become closely similar to vector 
mesons like the $\rho^0$, with which it shares quantum numbers, while 
high-virtuality ones remain of a perturbative character. The photon can
also interact in its simple unresolved form, even if this is the lesser
part of the total cross section at high energies. Together with 
Gerhard Schuler, a CERN fellow, a complete framework was developed for 
$\gamma\p$ physics e.g.\ at HERA \cite{Schuler:1993td} and $\gamma\gamma$ 
one e.g.\ at LEP~2 \cite{Schuler:1996en}. The full framework contains 
several process types and interaction scales already for real photons,
and when additionally the photons may be virtual the situation becomes 
even more complicated \cite{Friberg:2000ra}. 

A very special Higgs production channel at LHC is $\W\W$ or $\Z\Z$ gauge 
boson fusion, since the colour singlet nature of the exchanged particles 
seemingly would imply that the central rapidity range of the event would
be free of other particle production than that of the Higgs decay itself.
This is clearly not the case when MPIs are considered, where further 
interactions are likely to involve colour exchange and thus span strings
across the full rapidity range \cite{Dokshitzer:1991he}. There is still 
a small probability of no MPIs, ``the rapidity gap survival factor''
\cite{Bjorken:1992er}. Recently such MPI concepts were also used to 
describe the rate of jet production in diffractive events within the 
\textsc{Pythia} context \cite{Rasmussen:2015qgr}. 

Heavy flavours, from charm to top, involve their own sets of problems,
that have been studied from different angles over the years.
One interesting issue is the large charm vs. anticharm production 
asymmetries at fixed-target experiments, that can be understood in 
terms of the string topologies involved. In $\pi^-\p$ collisions with 
$\ubar\u \to \cbar\c$, for instance, the $\cbar$ is colour-connected 
to the $\pi^-$ beam remnant and pulled forwards by the string, while 
the $\c$ is connected to the $\p$ beam remnant and held back 
\cite{Norrbin:2000zc}. This gives rise to quite different momentum 
spectra, where some particle species thereby can take more momentum 
than the (anti)charm quark they come from. When strings are so short 
that they can collapse to a single particle, also the production rates 
become quite asymmetric. Asymmetries are smaller at higher energies,
and for $\b$ quarks, but not such that they can always be neglected in 
$CP$ violation studies, where they form a background.   

Before the top was found, alternative scenarios had to be considered. 
A ``light'' top would be long-lived enough that top hadrons had time to
form, while a heavy top would decay before that happens. With the latter
scenario confirmed, the issue immediately arises that there is no unique
set of colour singlet final-state particles that can be associated 
with the original colour triplet top quark. This leads to top mass 
uncertainties, e.g.\ from colour reconnection, that have been studied 
over the years \cite{Skands:2007zg,Argyropoulos:2014zoa}. But it also 
leads to nontrivial angular distributions, e.g.\ in
$\e^+\e^- \to \t \tbar \to \b \W^+ \bbar \W^-$ the emission of gluons 
with energies above the top width is essentially uncorrelated between
the $\b$ and $\bbar$, while soft gluons and nonperturbative hadronization
strongly correlate the two \cite{Khoze:1994fu}.   

Programs for matrix-element calculations have existed since long, 
but often as private code in no shape to be run by anybody else.
In the nineties this started to change. In the LEP~2 preparations,
some 15 different codes were available for four-fermion final states, 
some tailor-made and others general purpose \cite{Bardin:1997gc}. Since 
these did not address parton showers or hadronization, an ad hoc 
four-fermion interface was constructed to \textsc{Pythia}. But that
still left open what to do with the explosion of processes to be studied
at the LHC, and the increase of tools created to allow this study with 
improved precision. At the 2001 Les Houches meeting therefore a generic 
interface was developed between matrix element generators and 
shower+MPIs+hadronization programs \cite{Boos:2001cv}. 
This first Les Houches Accord consisted of two Fortran commonblocks.
One with beam and generation strategy information. The other 
with event-by-event listing of the particles involved in the hard
process, plus some weight and scale information. With the need to 
interoperate across languages, some years later the Les Houches Event 
Files offered a plain text alternative, initially carrying the same 
information \cite{Alwall:2006yp} but gradually expanded to cater to 
increasing needs.
   
One of the BSM physics areas not implemented in \textsc{Pythia} for a long 
time was Supersymmetry (SUSY). It has been popular with theorists and 
experimentalists over the years, more so than the BSM models that
actually were included. But it involves such an overwhelming set of 
parameters, particles, production processes and decay chains that 
realistic scenarios can become quite complex. 
It was also an area where most of the \textsc{Isajet} effort went in, 
so it was difficult to offer a competitive alternative. However, in 1994 
Stephen Mrenna, then a Caltech postdoc, informed me that he was 
implementing SUSY within the \textsc{Pythia} framework. This 
\textsc{SPythia} code was initially presented on its own 
\cite{Mrenna:1996hu}, but thereafter rapidly integrated into the 
regular distribution. Some years later Peter Skands, as master and PhD 
student, implemented lepton- and baryon-number-violating decays
\cite{Sjostrand:2002ip}, thereby further extending the framework. 
He also took the initiative to the SUSY Les Houches Accord (SLHA) 
\cite{Skands:2003cj}, that set a standard how couplings and particle 
properties could be transferred from spectrum calculators to event 
generators, and thereby greatly eased the task of setting up various 
SUSY scenarios. 

\section{The transition to C++}
\label{cplusplus}

The Fortran language had dominate scientific computing in particle 
phy\-sics since early days, and for many the assumption was that this
would continue, updated to more recent versions than the dominating 
Fortran~77 one. But in the nineties a growing group of
forerunners were busily advocating more modern languages, notably 
C++. The campaign succeeded, and eventually the CERN management decided 
not only that C++ would become the main language for the LHC era,
but also that the Fortran language would be phased out, to the extent 
that no Fortran compiler would be made available on CERN computers. 
This sent shock waves through the event generator community, needless 
to say, and we had to relate to a new reality in which we had been 
declared obsolete.

As it came to play out, the era of mainframe computing with proprietary 
expensive compilers was drawing to an end and, with the transition to
farms running free Linux/GCC software, the Fortran language continued 
to be available. This did not change the fact that the next generation
of event generator users in the experimental community would not be
taught Fortran, and that it was believed to be easier for students with
a C++ background to find jobs outside particle physics. 

So, in early 1998, Leif L\"onnblad and I began an intended 
project to convert at least most of the \textsc{Pythia} functionality
to C++ within three years. We came from different backgrounds, with
Leif being one of the early advocates of C++ and I having no previous
experience. The idea, however, was that Leif would do most of the work,
and we had ensured funding to this end. Unfortunately things did not 
work out as intended. Leif got involved in HERA studies and workshops,
slowing progress. A postdoc, who was quite keen to get involved, used 
the opportunity more as a way to explore C++ than to deliver working code.

There were also differences of philosophy. Firstly, Leif wanted to 
construct a sophisticated and powerful framework, while I was asking 
for a simple and easily understood structure, and we 
never managed to find a common middle ground. Secondly, after 
some years Leif made an agreement with the \textsc{Herwig} people that 
the then existing \textsc{Pythia}~7 base code would be renamed into 
\textsc{ThePEG} and form a common generator-neutral platform for both 
\textsc{Herwig++} \cite{Bahr:2008pv} and \textsc{Pythia} plugin modules, 
and \textsc{Sherpa} \cite{Gleisberg:2003xi} ones if that group joined. 
The advantage would be that a user would only need to learn one platform,
and that it would become simple to mix different physics models. This ran 
counter to my sentiments, that there is an intrinsic value in having 
completely independent codes to compare, to reduce the risks of common 
bugs, and that maybe not all combinations of models should be encouraged.

With progress stalled, the future looked bleak. Opportunity came in an
unexpected way, namely that Lund University underwent an economical
crisis, which also strongly affected our group. It therefore became 
feasible, and even encouraged, for me to go away for a longer period. 
The SFT group at CERN, in charge of developing common scientific
software such as \textsc{Root} \cite{Antcheva:2009zz} and \textsc{Geant}
\cite{Agostinelli:2002hh}, offered to take me in for three years of code 
writing. Part of the deal was that this would not be just a salvaging 
operation of \textsc{Pythia}~7 but a fresh new start.

Thus, 2004--08 (with an unintended break of a few months) I worked on 
the new \textsc{Pythia}~8 code close to full-time. It was a fortunate
timing, in between LEP and LHC, meaning less ``distraction'' from
physics questions and workshops than normally. Meanwhile Stephen and Peter 
largely took care of \textsc{Pythia}~6 and Tevatron support. 

The design philosophy was to keep the basic code as simple as possible. 
It was new code written from scratch, with few exceptions, but clearly 
inspired by strengths and shortcomings of the existing Fortran code. 
 
\begin{figure}[t]
\begin{center}
\includegraphics[width=0.8\linewidth]{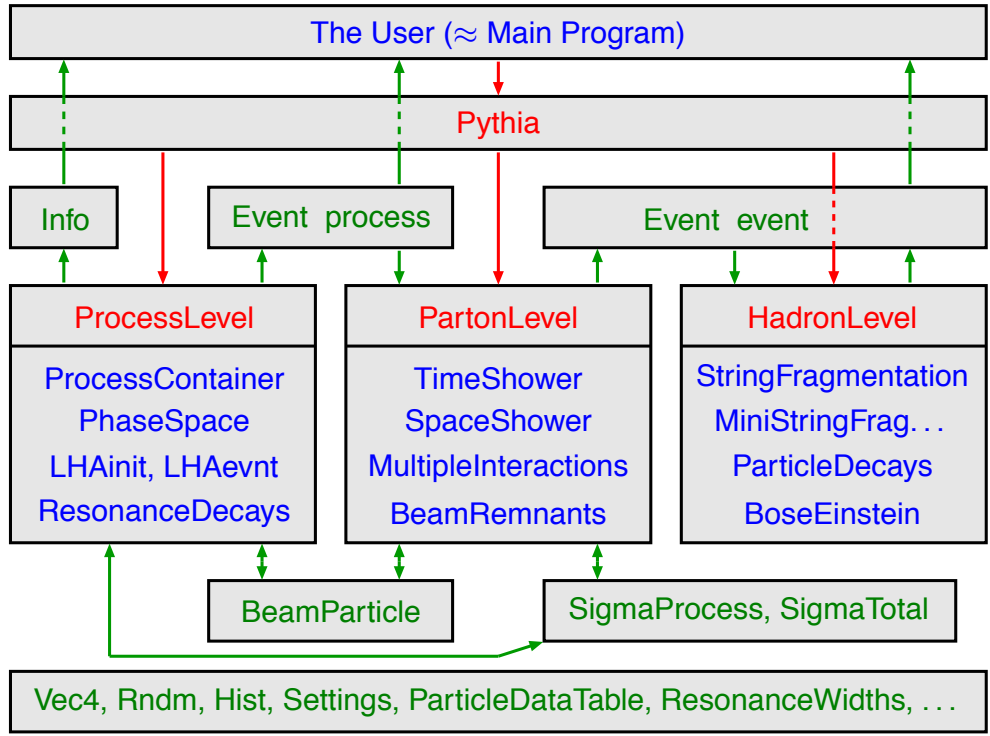}
\end{center}
\caption{\label{Fig:codestruct}Schematic view of the code structure in 
\textsc{Pythia}~8 as of 2007.}
\end{figure}

A schematic view of the relationship between the new classes is shown 
in Fig.~\ref{Fig:codestruct}. Some generic facilities, like four-vectors
and the settings and particle data databases, can be used just about
anywhere. Else most classes occur in a hierarchy, with \texttt{Pythia} 
on top, and below that three class sets representing main stages in
the evolution of an event, from hard perturbatively calculable processes
to soft nonperturbative ones.
\begin{itemize}
\item The \texttt{ProcessLevel} administrates the choice of a hard 
process by a combination of matrix element expressions and phase space 
selection. Most of the processes in \textsc{Pythia}~6 were taken over.
Input of externally generated events via the Les Houches accord,
an afterthought in \textsc{Pythia}~6, was here a standard option from 
the beginning. The operation of internal processes was even adjusted 
to conform. Specifically, whereas previously weak decay chains like
$\t \to \b \W^+ \to \b \q \qbar$ were only generated after
showers in previous steps had run their course, now they are generated 
already at the process level. The decays are stored as part of the 
\texttt{process} hard-process event record, from which information 
is then fetched in as needed later. For the few but important ``soft'' 
processes, such as inelastic nondiffractive events, the process-level 
stage is rather minimal, with the real action beginning later.
\item The \texttt{PartonLevel} traces the continued evolution down to
lower scales and higher partonic multiplicities, by including the effects
of parton showers on the hard process, by adding MPIs (in the figure 
called multiple interactions), and eventually by adding beam remnants.
Administratively, ISR, FSR and MPI are all three generated 
in an interleaved manner, in order of decreasing $\pT$ scales. 
The philosophy is that the harder activity sets the boundary conditions
for the softer one, irrespective of naive time ordering. 
Apart from the hard events discussed so far, also soft events can here 
be imparted with MPI activity, which may give ISR/FSR in its turn.
Other physics mechanisms, such as colour reconnection, can also occur 
at this stage.
\item The final \texttt{HadronLevel} step turns partons into hadrons
by string fragmentation, and takes care of subsequent decays of hadrons
and leptons. This step can also include other nonperturbative physics,
such as models for Bose-Einstein correlations.
\end{itemize}  
The main output of the generation is the \texttt{event} event record,
that documents both initial, intermediate and final particles generated
in the course of the three stages above.

Many generators are intended to be run by a set of input cards to a
single executable that, once built, does not need to be touched again. 
\textsc{Pythia}~8 can be run in this way, but it is not the only one.
Rather, it is the ability to use the program in different contexts that
lends real power. A set of example main programs is provided to 
illustrate this flexibility. The main program can contain quite 
sophisticated event analyses, making use of the power of having the 
full event history available, in combination with various analysis codes, 
notably a custom interface to  \texttt{FastJet} for jet clustering 
\cite{Cacciari:2011ma}. In addition to a multitude of settings that can 
affect the program execution, there is also the ability to insert external 
code at critical points along the generation chain, by plugins or user 
hooks. The former offer a way to replace some of the standard tasks by your
own code, notably for parton showers or random numbers. The user hooks
calls are interspersed with the normal code and allow optional extensions 
of it. Typically a hook is intended to allow for simple decisions, 
e.g.\ to veto some parton-shower emissions. As it happens, there has 
been a steady demand for more user hooks.   

The three-year period was sufficient to have \textsc{Pythia}~8.100 up 
and running, with about as much LHC physics functionality as the 
\textsc{Pythia}~6 code had. But time did not permit to convert 
everything, with $\gamma\p$, $\gamma\gamma$, SUSY and Technicolor physics 
being among the afflicted. And some obsolete options were scrapped 
altogether. 

\section{Current activities}
\label{current}

The \textsc{Pythia}~8 code has continued to expand after the original
release, and today more persons are involved in development work in
and around \textsc{Pythia} than ever before. Some main themes are 
described in the rest of this section.  

As perturbative calculational capabilities have progressed, it has become
possible to generate processes with more final-state particles, not only 
to leading order (LO) but also routinely to NLO,
and for some processes to even higher orders. Nevertheless, it is rarely 
if ever possible to use a pure ME approach to describe 
all perturbative activity down to the hadronization scale, as required 
to provide a consistent description, so one must still combine MEs with
parton showers. This field of activity is today 
called Match and Merge (M\&M), where match refers to providing a smooth 
transition from fixed-multiplicity MEs above some scale to showers 
below that, and merge to the consistent combination of different ME 
multiplicities. It can be argued that this has been the main theme of 
event-generator development work in the last twenty years. With some 
forerunners, like the already-mentioned exponentiation of matrix elements 
for the first emission, the real beginning was a 1999/2000 LEP~2 workshop.
Out of discussions there sprung the idea that 
LO MEs of different multiplicities could be combined consistently, 
provided that they were corrected with Sudakov form factors 
describing the no-emission virtual corrections.
Two related approaches were developed, the CKKW one \cite{Catani:2001cc}, 
where the Sudakovs are obtained from the analytical expression, and the 
CKKW-L one \cite{Lonnblad:2001iq}, where explicit ``trial showers'' are 
used to generate the Sudakovs. Today the latter approach has become the 
norm, but has branched out in several variants. 
     
M\&M methods were not part of the distributed \textsc{Pythia}~6 code, 
but a number of studies were still done using different add-on codes 
\cite{Mrenna:2003if, Alwall:2007fs}, e.g. as part of \textsc{MadGraph} 
\cite{Alwall:2007st}. In \textsc{Pythia}~8 several such methods are 
part of the distribution, but usually as plugins rather than in the 
core library. However, in 2011 Leif L\"onnblad together with then-student
Stefan Prestel began to develop and implement M\&M methods in the 
\textsc{Pythia}~8 core, addressing detailed issues such as how 
interleaving of showers with MPIs should be taken into account
\cite{Lonnblad:2011xx}. (Which involved the creation of a whole new 
\texttt{PartonLevel} instance to run trial showers and MPIs.)
The development extended to include also 
usage of NLO MEs \cite{Lonnblad:2012ix,Lonnblad:2012ng} in a few 
different approaches. This code is still actively developed to cover 
more kinds of processes and higher multiplicities, among others.
Combined with the ones previously mentioned, there is currently 
approximately ten different M\&M schemes available to play with. 
In addition, M\&M in external ME programs like \textsc{MadGraph} and 
the \textsc{PowHeg~Box} \cite{Alioli:2010xd} are also commonly used.

With the ME part of the story increasingly well modelled, it is natural
also to develop the shower part to have higher precision. Early on
\textsc{Pythia}~8 was opened up to allow external shower programs
to be linked in as a replacement for the native ones, and over the 
years this has been used for some studies. Of special interest are
the \textsc{Vincia} \cite{Giele:2007di} and \textsc{Dire}  
\cite{Hoche:2015sya} projects. Both have the ambition to raise showers 
to next-to-leading or even next-to-next-to leading logarithmic precision. 
\textsc{Vincia}, by Peter Skands and coworkers, is the older of the two. 
Over the years it has been used to test out a number of new approaches, 
such as smooth ordering, sector showers, antenna showers (for FSR+ISR), 
helicity-dependent showers, and iterated ME corrections. \textsc{Dire} 
is a rather unusual project, in that the same shower algorithm has been 
coded twice, once for \textsc{Sherpa} by Stefan H\"oche and once for 
\textsc{Pythia} by Stefan Prestel, so as to reduce the risk of bugs. 
Unlike most other codes it gives variable and even negative weights 
already at leading log. 

Both \textsc{Vincia} and \textsc{Dire} are large programs in their own 
right, not only by the core evolution code itself, but by the environment 
of matrix elements, splitting libraries, M\&M machineries and more. 
The advantage of having been freestanding is that development and new 
releases has been decoupled from the \textsc{Pythia} ones. The disadvantage 
is that the threshold for users to try out these showers is higher. 
For the upcoming \textsc{Pythia}~8.3 release, therefore, they will 
become part of this distribution. They will still continue to be 
vigorously developed as identifiable subpackages. 

Also the default \textsc{Pythia} parton showers have continued to evolve 
over the years, e.g.\ by the optional emission of weak gauge bosons
\cite{Christiansen:2014kba} and by a framework for showers in various 
Hidden Valley scenarios \cite{Carloni:2011kk}. The latter also includes 
hadronization in the secluded sector, and decay back into the visible 
sector. Similarly, production of long-lived coloured particles, such as
squarks and gluinos in some scenarios, combine showers and hadronization.
First these particles can radiate, then hadronize into so-called 
$R$-hadrons that propagate some distance, then decay in a process that 
again will involve showers and a new hadronization step.

Traditionally the emphasis of \textsc{Pythia} has been on incoming 
hadron or lepton beams, including somewhat special cases like photons 
and Pomerons. But already in the mid-eighties a model \textsc{Fritiof}
\cite{Andersson:1986gw} and related program \cite{NilssonAlmqvist:1986rx}
was developed for hadronic reactions, with generalization to heavy-ion 
collisions. In it  nucleons could get an increasingly excited mass by 
successive collisions. The resulting states would undergo \textsc{Jetset} 
string fragmentation after the two nuclei had passed through each other. 
The model was quite successful for fixed-target energies, but perturbative 
parton--parton scatterings were difficult to include, and therefore it 
could not reliably be used at higher energies. A few years ago heavy-ion 
activities were started up again, with the \textsc{Angantyr} model 
\cite{Bierlich:2018xfw}. It is based on the conventional MPI model
for $\p\p$ collisions, and allows for a nucleon to undergo successive
collisions, where one is of the conventional $\p\p$ type and the rest 
of them can be viewed as a Pomeron
taken out of the nucleon colliding with a nucleon from the other nucleus. 
This is similar to how the beam-remnant machinery handles MPIs in 
ordinary $\p\p$ collisions, such that not all strings stretch out all
the way to the edge of the allowed rapidity range but stay more centrally.   
The model also includes shove, that two strings can repel each other 
and thereby give collective flow and azimuthal anisotropies, and ropes,
that two strings can overlap to give a higher string tension that favours 
the production of more strangeness. The objective is to see how far it
is possible to go with the description of heavy-ion collisions without
invoking the existence of a quark--gluon plasma. Thus the development of 
\textsc{Angantyr} and related aspects will be a central undertaking 
for the coming years. 

A more recent project has been to pin down the space--time structure 
of the hadronization process in $\p\p$ collisions 
\cite{Ferreres-Sole:2018vgo}. This is a first step towards modelling 
hadronic rescattering, initially for $\p\p$ but later possibly also for 
heavy-ion collisions. An important component is the modelling of 
hadronic collisions from threshold energies upwards, which is not possible
currently, and where the collision energy and hadron types can be changed 
flexibly. Such a machinery could also find other applications, e.g.\ for 
cosmic-ray cascades in the atmosphere.   

Other physics studies are ongoing within the \textsc{Pythia} context, like 
photoproduction and $\gamma\gamma$ physics, diffraction and total cross 
sections, neutrino interactions, Dark Matter processes and other BSM 
physics, coalescence processes for deuterium production, production of 
charmonium and bottomonium in showers, new possibilities for DPS studies,
and more. Projects like that will always show up and hopefully leave their 
imprint in the code available for users.

Most of the \textsc{Pythia} model components involve free parameters 
that have to be determined by comparisons with data. The total number 
of parameters is quite large and typically these are correlated in 
nontrivial ways. Therefore, from the early days onwards, the production 
of internally consistent tunes has been a recurring activity. This is a 
task performed both by \textsc{Pythia} members and by the various 
experimental collaborations. Unfortunately the latter kind of tunes tend 
to be restricted to data from the own experiment, and sometimes only 
to a subset, e.g.\ either to minimum-bias or to high-$\pT$ jet data. 
That way agreement can be improved for the tasks at hand, at the
expense of worse agreement elsewhere. Global detector-agnostic tunes 
therefore have a relevant role to play. Current default values are based 
on the Monash~2013 global tune \cite{Skands:2014pea}, which also has 
served as a starting point for other tunes.

In addition to physics modelling and studies, there is also a 
significant upgrade in the works, namely that the upcoming 
\textsc{Pythia}~8.3 release will be based on C++11 rather than on 
C++98. This means e.g.\ that smart pointers, container loops and
function wrappers will come into use, initially at a few places in 
the existing code but likely to be more common in new code. Some of 
the other new C++11 features we have decided not to use, so as not to 
introduce unnecessary complication and make code less transparent. 

There are also other changes to the code, some of which are unrelated 
to the language upgrade, but suitable to implement when backwards
compatibility is anyway (mildly) broken. One such is a new \texttt{InfoHub}
class that includes several other service classes, thereby reducing
the number of arguments needed to pass around (pointers to) other classes. 
This is combined with a new \texttt{PhysicsBase} base class, from which  
several of the physics classes are derived, with automatic import of
\texttt{InfoHub}, and an option to set up methods that are called before 
and after each event has been generated. The intention is also to 
improve the parton-shower interface, to streamline the \textsc{Vincia} 
and \textsc{Dire} integration. There will be a new interface for various
string interaction models.

The XML/HTML documentation has expanded significantly since the 
original 8.1 release, and the separate pages are now being reordered 
for better cohesion. A searchable index of the example main programs 
is added. In the future the pages should gradually be expanded with more 
descriptions of the physics involved, eventually leading up to a new
publishable long 8.3 manual. The PHP version is discontinued, since it 
would require double work to maintain within the intended new structure.

\section{Summary and outlook}
\label{summary}

As we have seen, the \textsc{Jetset/Pythia} project has expanded, 
from a humble project to study some simple distributions within 
a single jet, to a code with intentions to be relevant for essentially 
all areas of high-energy particle physics.
 
\begin{figure}[t]
\begin{center}
\includegraphics[width=0.8\linewidth]{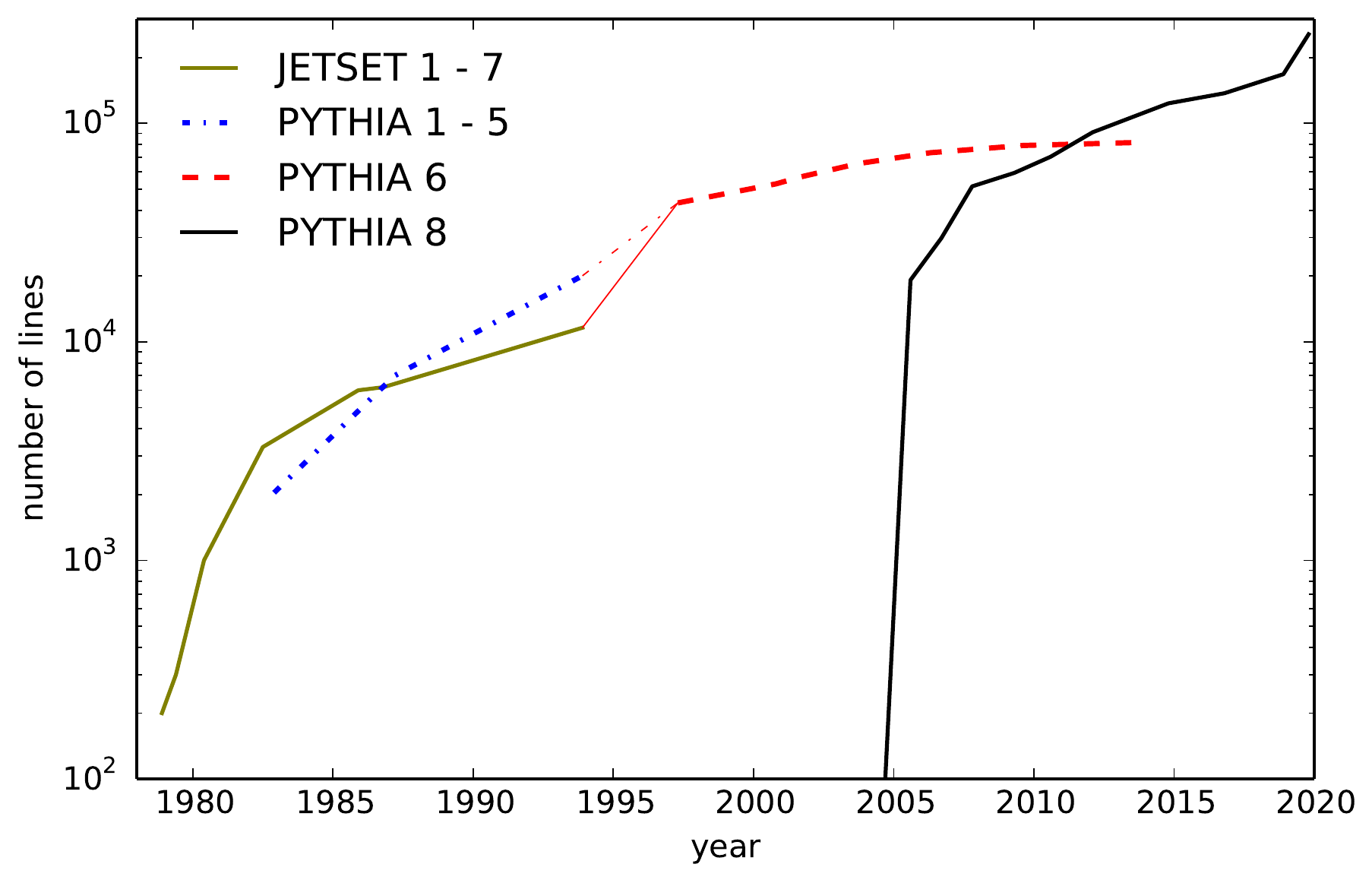}
\end{center}
\caption{\label{Fig:programsize}Number of lines in the program codes 
as a function of time. Snapshots in time are connected by straight 
lines. Thin lines around 1995 mark the merger of \textsc{Pythia} and
\textsc{Jetset}.}
\end{figure}

As a natural consequence, the \textsc{Pythia} code size has been 
steadily increasing over the years; obviously with \textsc{Pythia}~8 
starting over from scratch but then rapidly overtaking \textsc{Pythia}~6. 
The discovery of the string effect was based on a code with a total 
size of 1\,000 lines, while projections for the 8.3 release hover 
around 250\,000 ones. A big jump from the current 8.2 size is 
explained by the inclusion of the \textsc{Vincia} and \textsc{Dire} 
shower codes. This evolution is shown in Fig.~\ref{Fig:programsize}. 
It is largely based on what the \texttt{wc} command gives, which 
include both comment and blank lines, and is only for the core program, 
which for the C++ versions are the \texttt{include/Pythia8} and 
\texttt{src} subdirectories. The actual \textsc{Pythia}~8 code 
distribution is larger, with example programs, parton distribution 
function data files, manual pages, and more. So far the original
organizational structure of the code has been possible to extend 
gradually and reasonably smoothly, but this may not always be the case.

The current and previous versions of the \textsc{Pythia} code, along with
auxiliary documentation and files, and relevant presentations by team 
members, can be found at the \textsc{Pythia} webpage
\begin{center}
\texttt{http://home.thep.lu.se/Pythia}
\end{center} 

In one respect \textsc{Pythia}~8 is still trailing \textsc{Pythia}~6,
namely in the size of the manual. The current XML/HTML-based manual
does document all settings and all user-accessible methods, but is
rather brief in the physics descriptions. Most is documented in
separate physics articles, of course, but from these it is not always
possible to get a coherent view over which different ideas have been 
implemented and combined how. One of the projects for the future 
is to improve the physics documentation, both on the separate 
HTML web pages and as a combined overview.  

The size of the development team has fluctuated over the years, 
but is currently rising at a steady pace. It remains quite Lund-centric,
with most members being either former (Christian Bierlich, 
Leif L\"onnblad, Stefan Prestel, Peter Skands, myself) or current
(Christine Rasmussen) PhD students. Others have been recruited as 
postdocs (Ilkka Helenius) or short-term students (Nishita Desai,
Philip Ilten) within the MCnet collaboration of event generator authors.
Only one person (Stephen Mrenna) came in directly from the outside.
Note that the number of members does not translate directly into 
man-years of \textsc{Pythia} development, since for many this is
one activity among others. 

In the LEP and early LHC days I was alone to answer all questions people 
might have, which took quite a significant chunk out of my daily work. 
The bigger size is quite helpful in this respect, both for sharing the 
basic questions and for assigning the more specialized ones according
to our respective areas of expertise. By chance there is also a good 
spread of LHC contacts, in that Stephen is a CMS member and Phil an LHCb 
one, while Stefan is an ATLAS ``Analysis Consultant \& Expert'' and 
Christian is closely working together with ALICE people.

The organizational structure has not quite kept pace with the expanding
team. By tradition things are run informally, and people mainly contribute 
to their areas of interest (or those of their supervisor). The ambition
of monthly Skype/Vidyo meetings has been limping. A 2019 week-long meeting 
at Monash University (Melbourne, Australia) was the first-ever organized
in-person chance to discuss the near- and far-term development of the
program. The former is reflected in the strategy for a transition from
C++98 to C++11, now well along the way, and the subsequent release of 
\textsc{Pythia}~8.3. The latter involve plans for the next manual but 
also less conclusive discussions on the future organizational structure, 
including issues such as having internal reviewers before new code is 
added to the public version. The key message, however, is that the 
\textsc{Pythia} collaboration intends to be in there for the long haul.

We have already touched on areas of ongoing activities, 
such as heavy-ion physics, improved parton showers, or rescattering.
More areas are likely to arise in the future, unknown now exactly which,
but partly related to future directions in experimental physics. 
FCC-hh is a natural evolution from LHC and FCC-ee from LEP, so can 
already be run with reasonable expectations of reliably foretelling
what to expect. For CLIC the challenge is the large background from
beamstrahlung interactions, which partly can be modelled already now
but may require further development. More work would definitely be 
needed for the EIC and FCC-eh, both to consistently model the transition 
region between photoproduction and DIS and to handle nuclear effects.
 
Finally, it is important to note that \textsc{Pythia} has played
a dual role throughout its history. On the one hand, we have striven for 
increased precision/accuracy in predicting/describing experimental data,
thereby offering an indispensable tool for the experimental community.
But, on the other hand, it has also been a way to explore new ideas,
say in topics that may be beyond perturbative control, and in the end 
these have often been as important for the advancement of the field.
The hope and aim is that this dual nature will carry \textsc{Pythia} on
into the future. 

\section*{Acknowledgements}

The progress outlined in this article would not have been possible
without the collaboration of a large number of persons, including but
not limited to past and present \textsc{Pythia} collaboration members,
who have contributed both ideas and code. Apologies that not all such 
contributions are documented here.

Work supported in part by the Swedish Research Council, contract number
2016-05996, and in part by the MCnetITN3 H2020 Marie Curie Innovative 
Training Network, grant agreement 722104.
This project has also received funding from the European Research
Council (ERC) under the European Union's Horizon 2020 research
and innovation programme, grant agreement No 668679.

\end{document}